\documentclass[sigconf,natbib=true,screen=true]{acmart}

\usepackage{color}
\usepackage{bbm}
\usepackage{multirow}
\usepackage[inline]{enumitem}
\usepackage{graphicx}
\usepackage{subcaption}
\usepackage{sistyle}
\usepackage[ruled]{algorithm2e}
\usepackage{booktabs}
\usepackage{caption}
\SIthousandsep{,}
\usepackage{makecell}
\usepackage{amsmath}
\usepackage[english]{babel}
\usepackage{hyperref}
\usepackage{array} 
\usepackage{graphicx}
\usepackage{algorithmic}
\usepackage{tikz}
\usepackage{wrapfig}
\usepackage{acronym}
\newcommand{\heading}[1]{\vspace*{.75mm}\noindent\textbf{#1.}}

\AtBeginDocument{%
  \providecommand\BibTeX{{%
    \normalfont B\kern-0.5em{\scshape i\kern-0.25em b}\kern-0.8em\TeX}}}

\makeatletter
\g@addto@macro\normalsize{%
  \abovedisplayskip 3pt plus1pt 
  \belowdisplayskip 3pt plus1pt
  \abovedisplayshortskip  0pt plus1pt%
  \belowdisplayshortskip  0pt plus1pt
}
\makeatother

\acrodef{CV}{computer vision}
\acrodef{IR}{information retrieval}
\acrodef{LLM}{large language model}
\acrodef{MDP}{Markov decision process}
\acrodef{NLP}{natural language processing}
\acrodef{NRM}{neural ranking model}
\acrodef{RL}{reinforcement learning}
\acrodef{RL-MARA}{Multi-grAnular Ranking Attack}
\acrodef{MoE}{mixture-of-experts}

\copyrightyear{2024}
\acmYear{2024}
\setcopyright{rightsretained}
\acmConference[SIGIR '24]{Proceedings of the 47th International ACM SIGIR Conference on Research and Development in Information Retrieval}{July 14--18, 2024}{Washington, DC, USA}
\acmBooktitle{Proceedings of the 47th International ACM SIGIR Conference on Research and Development in Information Retrieval (SIGIR '24), July 14--18, 2024, Washington, DC, USA}
\acmDOI{10.1145/3626772.3661380}
\acmISBN{979-8-4007-0431-4/24/07}

\makeatletter
\gdef\@copyrightpermission{
  \begin{minipage}{0.3\columnwidth}
   \href{https://creativecommons.org/licenses/by/4.0/}{\includegraphics[width=0.90\textwidth]{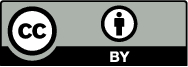}}
  \end{minipage}\hfill
  \vspace{5pt}
}
\makeatother



\author{Yu-An Liu}
\orcid{0000-0002-9125-5097}
\author{Ruqing Zhang}
\orcid{0000-0003-4294-2541}
\affiliation{
	\institution{CAS Key Lab of Network Data Science and Technology, ICT, CAS}
	\institution{University of Chinese Academy of Sciences}
	\city{Beijing}
	\country{China}
}
\email{{liuyuan21b,zhangruqing}@ict.ac.cn}

\author{Jiafeng Guo}
\orcid{0000-0002-9509-8674}
\affiliation{
	\institution{CAS Key Lab of Network Data Science and Technology, ICT, CAS}
	\institution{University of Chinese Academy of Sciences}
	\city{Beijing}
	\country{China}
}
\email{guojiafeng@ict.ac.cn}

\author{Maarten de Rijke}
\orcid{0000-0002-1086-0202}
\affiliation{
 \institution{University of Amsterdam}
 \city{Amsterdam}
 \country{The Netherlands}
}
\email{m.derijke@uva.nl}

\begin{document}

\title[Robust Information Retrieval]{Robust Information Retrieval}

\begin{abstract}
Beyond effectiveness, the robustness of an information retrieval (IR) system is increasingly attracting attention. When deployed, a critical technology such as IR should not only deliver strong performance on average but also have the ability to handle a variety of exceptional situations. In recent years, research into the robustness of IR has seen significant growth, with numerous researchers offering extensive analyses and proposing myriad strategies to address robustness challenges. In this tutorial, we first provide background information covering the basics and a taxonomy of robustness in IR. Then, we examine adversarial robustness and out-of-distribution (OOD) robustness within IR-specific contexts, extensively reviewing recent progress in methods to enhance robustness. The tutorial concludes with a discussion on the robustness of IR in the context of large language models (LLMs), highlighting ongoing challenges and promising directions for future research. This tutorial aims to generate broader attention to robustness issues in IR, facilitate an understanding of the relevant literature, and lower the barrier to entry for interested researchers and practitioners.
\end{abstract}

\maketitle

\section{Motivation}
Information retrieval (IR) systems are an important way for people to access information.
In recent years, with the development of deep learning, deep neural networks have begun to be applied in IR systems \cite{guo2016deep,chen2017efficient,liu2017cascade}, achieving remarkable effectiveness.
However, beyond their effectiveness, these neural IR models also inherit the inherent \emph{robustness flaws} of neural networks \cite{wu2022prada,wang2022bert,thakur2021beir}. 
This poses a hindrance to their widespread application in the real world.

In the past few years, the issue of the robustness of IR has received wide attention, e.g., \citet{wu2022neural} analyzed the robustness of neural ranking models (NRMs), and a perspective paper on competitive search \cite{kurland2022competitive} discussed adversarial environments in search engines. 
Since then, there has been a lot of work that focuses on different robustness aspects in IR, such as adversarial robustness \cite{wu2022prada,liu2022order,liu2023black,liu2024multi}, out-of-distribution (OOD) robustness \cite{thakur2021beir,chen2022towards}, performance variance \cite{wu2022neural}, robustness under long-tailed data \cite{garigliotti2019unsupervised}, and on the corresponding improvement options.
Today, the research community can effectively scrutinize IR models leading to more robust and reliable IR systems. 

To ensure the quality of the tutorial, we will focus on the two most widely studied types of robustness issues, namely \emph{adversarial robustness} and \emph{OOD robustness}. 
There are many analyses and suggestions for improvement around these two robustness issues, but it has not yet been systematically organized.
Through this tutorial, we aim to summarize and review the progress of robust IR to attract attention and promote widespread development.


\section{Objectives}
\heading{1. Introduction}
We start by reminding our audience of the required background and introducing the motivation and scope of the robustness issue in IR in our tutorial.

\heading{2. Preliminaries}
In IR, robustness signifies an IR system's consistent performance and resilience against diverse unexpected situations. 
There is a large volume of work that covers many aspects of IR robustness, e.g., 
\begin{enumerate*}[label=(\roman*)]
\item \textit{Adversarial robustness}  \cite{wu2022prada, liu2023black}, which focuses on the ability of the IR model to defend against malicious adversarial attacks aimed at manipulating rankings;
\item \textit{OOD robustness}  \cite{thakur2021beir,wu2022neural}, which measures the performance of an IR model on unseen queries and documents from different distributions of the training dataset;
\item \textit{Performance variance}  \cite{wu2022neural}, which emphasizes the worst-case performance across different individual queries under the independent and identically distributed (IID) data; and
\item \textit{Robustness under long-tailed data} \cite{garigliotti2019unsupervised}, which refers to the capacity to effectively handle and retrieve relevant information from less common, infrequently occurring queries or documents.
\end{enumerate*}

In this tutorial, we focus on adversarial robustness and OOD robustness, which have received the most attention.
Interest in adversarial robustness stems largely from the widespread practice of search engine optimization (SEO) \cite{castillo2011adversarial}.
Concerns about OOD robustness are primarily due to the need for adaptation across diverse and complex real-world scenarios. 
Moreover, as large language models (LLMs) are being integrated into IR, new robustness challenges emerge; LLMs also offer novel opportunities for enhancing the robustness of IR systems.

Building on these preliminaries, we will cover adversarial robustness, OOD robustness, and robust IR in the age of LLMs.

\heading{3. Adversarial robustness}
The web is a competitive search environment, which can lead to the emergence of SEO, in turn causing a decline in the content quality of search engines \cite{kurland2022competitive,castillo2011adversarial}.
With the gradual rise of SEO, traditional web spamming \cite{gyongyi2005web} started to become an effective way to attack IR systems.
However, this approach based on keyword stacking is easily detected by statistical-based spamming detection methods \cite{zhou2009osd}.

\textit{Adversarial attacks}. 
In order to exploit the vulnerability of neural IR models, many research works have simulated real black-hat SEO scenarios and proposed a lot of adversarial attack methods.
\begin{enumerate*}[label=(\roman*)]
\item First, we introduce the differences between attacks in IR and CV/NLP, including task scenarios and attack targets;
\item Then, we present adversarial retrieval attacks \cite{liu2023black,zhong2023poisoning,long2024backdoor,lin2023mawseo,boucher2023boosting} against the first-stage retrieval models, including the task definition and evaluation.
Current retrieval attack methods mainly include corpus poison attacks \cite{liu2023black,zhong2023poisoning,lin2023mawseo}, backdoor attacks 
     \cite{long2024backdoor}, and encoding attacks \cite{boucher2023boosting}; and
\item Finally, we introduce adversarial ranking attacks \cite{wang2022bert,wu2022prada,liu2022order,liu2023topic,chen2023towards,song2020adversarial,liu2024multi} against NRMs with task definitions and evaluation setups.
These include word substitution attacks \cite{wang2022bert,wu2022prada,liu2023topic}, trigger attacks \cite{liu2023topic,liu2022order,song2020adversarial}, prompt attacks \cite{chen2023towards,parry2024analyzing}, and multi-granular attacks \cite{liu2024multi}.
\end{enumerate*}

\textit{Adversarial defense}. 
To cope with adversarial attacks, research has proposed a series of adversarial defense methods to enhance the robustness of IR models.
\begin{enumerate*}[label=(\roman*)]
\item We introduce the objective and evaluation of IR defense tasks.
Based on these defense principles, adversarial defense methods in IR can be classified as attack detection, empirical defense, and certified robustness;
\item We turn to attack detection, which includes perplexity-based, linguistic-based, and learning-based detection \cite{chen2023defense};
\item We present empirical defenses, which encompass data augmentation \cite{chen2023dealing}, traditional adversarial training \cite{park2019adversarial,lupart2023study}, and theory-guided adversarial training \cite{liu2024perturbation}; and
\item We introduce the certified robustness in IR \cite{wu2022certified}.
\end{enumerate*}

\heading{4. Out-of-distribution robustness}
In real-world scenarios, search engines are in an ever-changing data environment, and new data are often not IID with the training data. 
Therefore, the ability to generalize to OOD data or not is the basis for the evaluation of IR systems in terms of OOD robustness \cite{wu2022neural}.

\textit{OOD generalizability on unseen documents}. 
In IR, the OOD robustness scenarios that have been examined can be categorized into unseen documents and unseen queries.
The unseen documents scenario may be caused by adaptation to new corpus \cite{thakur2021beir} or by incrementation of original corpus \cite{cai2023l2r}.
\begin{enumerate*}[label=(\roman*)]
\item Adaptation to new corpus usually refers to the phenomenon that the corpus on which an IR model is trained is not in the same domain as the corpus on which it is tested. 
Due to the overhead of retraining, the performance of the model on the new domain needs to be guaranteed under zero/few-shot scenario, which is usually solved by data augmentation \cite{Oh2023Data,Chen2023Cross-domain,Anaya-Isaza2022Data}, domain modeling \cite{zhan2022disentangled,yu2022coco}, architectural
modification \cite{huebscher2022zero,formal2022distillation}, scaling up the model capacity \cite{ni2022large,lu2022ernie}; and
\item Incrementation of original corpus refers to the scenario where new documents are continuously added to the corpus with potential distribution drift.
In this situation, the IR model should effectively adapt to the evolving distribution with the unlabeled new-coming data, which is usually solved by continual learning~\cite{chen2023continual,cai2023l2r}.
\end{enumerate*}

\textit{OOD generalizability on unseen queries}. 
Unseen queries concern query variations \cite{zhuang2021dealing} and unseen query types \cite{wu2022neural}.
\begin{enumerate*}[label=(\roman*)]
\item The query variations are usually different expressions of the same information need \cite{zhuang2021dealing, penha2022evaluating,liu2023robustness}  which may impact the effectiveness of IR models.
Many noise-resistant approaches such as self-teaching
methods \cite{chen2022towards,zhuang2022characterbert,zhuang2023typos}, contrastive learning methods \cite{zhuang2021dealing,liu2023mirs,ma2021contrastive,sidiropoulos2024improving}, hybrid methods \cite{tasawong2023typo,campos2023noise,pan2023towards} have been proposed for neural IR models; and
\item Unseen query types refer to the unfamiliar query type with new query intents \cite{wu2022neural}.
Domain regularization \cite{cohen2018cross} is effective for dealing with new query types.
\end{enumerate*}

\heading{5. Robust IR in the age of LLMs}
\begin{enumerate*}[label=(\roman*)]
\item We first discuss the potential robustness challenges with applications of LLMs in IR, such as retrieval augmentation \cite{izacard2023atlas,fan2024right,ni2024llms}, and LLMs for ranking \cite{sun2023chatgpt,zhang2024large}; and
\item Then, we will discuss how LLMs can be used to enhance the robustness of IR systems.
\end{enumerate*}
These explorations will inspire many novel attempts in this area.

\heading{6. Conclusions and future directions}
We conclude our tutorial by discussing several important questions and future directions, including
\begin{enumerate*}[label=(\roman*)]
\item There is a diverse focus on the robustness of IR models from multiple perspectives. 
Establishing a unified benchmark of analysis to systematically analyze the robustness of all aspects of existing models.
\item For adversarial robustness, existing work on adversarial attacks focuses on specific stages (first-stage retrieval or re-ranking) \cite{wu2022prada,liu2023black} in IR systems.
Customizing adversarial examples to make them effective for all stages is challenging.
Therefore, one potential future direction is to explore how we can design a general unified attack method that can cater to every IR stage.
\item For OOD robustness, the main limitation of existing work is the difficulty of seeing enough diverse domain data in advance, leading to insufficient transfer capabilities of the model.
Using the generation capabilities of LLMs to synthesize corpora for adaptation domains seems to be a promising direction.
\end{enumerate*}

\section{RELEVANCE TO THE IR COMMUNITY}
In recent years, a considerable number of tutorials focusing on the topic of robustness have emerged across disciplines within computer science.
In KDD'21 \cite{datta2021machine}, CVPR'21 \cite{chen2021practical}, and AAAI'22, there were tutorials on robustness for AI and computer vision.
In EMNLP'21 \cite{chang2021robustness} and EMNLP'23 \cite{xu2023security}, there were tutorials on robustness and security challenges in NLP.
The focus of these tutorials was not on search tasks and models.

Search and ranking is a core theme at SIGIR. 
Evaluation, another core theme at SIGIR, encompasses multiple critical criteria beyond effectiveness for evaluating an IR system.
Robust information retrieval aligns well with these core themes.
Recently, robust information retrieval has gained considerable attention as more and more work is now devoted to analyzing and improving the robustness of information retrieval systems \cite{liu2022order,wu2022neural,goren2020ranking,thakur2021beir}.
Our tutorial will describe recent advances in robust information retrieval and shed light on future research directions. 
It would benefit the community and help to encourage further research into robust IR.

\section{FORMAT AND DETAILED SCHEDULE}
A detailed schedule for our proposed half-day tutorial (three hours plus breaks), which is aimed at delivering a high-quality presentation within the selected time frame, is as follows:

\begin{enumerate}[label=\arabic*., leftmargin=*]
  \item \textbf{Introduction} (15 minutes)
    \begin{itemize}
      \item Introduction to robust IR: motivation and scope
      \item Tutorial overview
    \end{itemize}
  \item \textbf{Preliminaries} (20 minutes) 
    \begin{itemize}
      \item Definition of robustness in IR
      \item Taxonomy of robustness in IR
    \end{itemize}
  \item \textbf{Adversarial Robustness} (50 minutes) 
    \begin{itemize}
      \item Traditional Web spamming
      \item Adversarial attacks
        \begin{enumerate}[label=-]
          \item Comparison: IR attacks vs. CV/NLP attacks
          \item Retrieval attacks: definition, evaluation, method, etc.
          \item Ranking attacks: definition, evaluation, method, etc.
        \end{enumerate}
      \item Adversarial defense
        \begin{enumerate}[label=-]
          \item IR defense tasks: objective \& evaluation
          \item Empirical defense: adversarial training, detection, etc.
          \item Theoretical defense: certified defense, etc.
        \end{enumerate}
    \end{itemize}
  \item \textbf{Out-of-distribution Robustness} (45 minutes) 
    \begin{itemize}
      \item OOD generalizability in IR
      \item OOD generalizability on unforeseen corpus
        \begin{enumerate}[label=-]
          \item Definition \& evaluation
          \item Adaptation to new corpus
          \item Incrementation of original corpus
        \end{enumerate}
      \item OOD generalizability on unforeseen queries
        \begin{enumerate}[label=-]
          \item Definition \& evaluation
          \item Query variation
          \item Unseen query type
        \end{enumerate}
    \end{itemize}
  \item \textbf{Robust IR in the Age of LLMs} (20 minutes)
        \begin{enumerate}[label=-]
          \item New challenges to IR robustness from LLMs
          \item New solutions for IR robustness via LLMs
        \end{enumerate}
  \item \textbf{Challenges and Future Directions} (20 minutes)
  \item \textbf{QA Session} (10 minutes)
\end{enumerate}

\section{TUTORIAL MATERIALS}
We plan to make all teaching materials available online for attendees, including: 
\begin{enumerate*}[label=(\roman*)]
\item Slides: The slides will be made publicly available.
\item Annotated bibliography: This compilation will contain references listing all works discussed in the tutorial, serving as a valuable
resource for further study.
\item Reading list: We will provide a reading list with a compendium of existing work, open-source code libraries, and datasets relevant to the work discussed in the tutorial.
\end{enumerate*}
We intend to ensure that all instructional materials are available online.\footnote{\url{https://robust-information-retrieval.github.io}}
Moreover, we grant permission to include slides and video recordings in the ACM anthology.

\begin{acks}
This work was funded by the National Key Research and Development Program of China under Grants No. 2023YFA1011602, the Strategic Priority Research Program of the CAS under Grants No. XDB0680102,  the project under Grants No. JCKY2022130C039, and the Lenovo-CAS Joint Lab Youth Scientist Project. 
This work was also (partially) funded by 
the Hybrid Intelligence Center, a 10-year program funded by the Dutch Ministry of Education, Culture and Science through the Netherlands Organisation for Scientific Research, \url{https://hybrid-intelligence-centre.nl}, 
project LESSEN with project number NWA.1389.20.183 of the research program NWA ORC 2020/21, which is (partly) financed by the Dutch Research Council (NWO),
project ROBUST with project number KICH3.LTP.\-20.006, which is (partly) financed by the Dutch Research Council (NWO), DPG Media, RTL, and the Dutch Ministry of Economic Affairs and Climate Policy (EZK) under the program LTP KIC 2020-2023,
and
the FINDHR (Fairness and Intersectional Non-Discrimi\-nation in Human Recommendation) project that received funding from the European Union’s Horizon Europe research and innovation program under grant agreement No 101070212.

All content represents the opinion of the authors, which is not necessarily shared or endorsed by their respective employers and/or sponsors.
\end{acks}

\bibliographystyle{ACM-Reference-Format}
\balance
\bibliography{references}

\end{document}